\PassOptionsToPackage{table,xcdraw}{xcolor}
\documentclass[conference]{IEEEtran}
\IEEEoverridecommandlockouts

\usepackage{xcolor}

\usepackage{cite}

\ifCLASSINFOpdf
  \usepackage[pdftex]{graphicx}
\else
\fi
\usepackage{url}
\usepackage{hyperref}
\usepackage{booktabs}
\usepackage[misc]{ifsym}

\usepackage{listings}
\usepackage{bbding}
\usepackage{graphicx}
\usepackage[table,xcdraw]{xcolor}

\definecolor{darkgreen}{rgb}{0, 0.44, 0.23}
\definecolor{lightgreen}{rgb}{0.25, 0.63, 0.4375}
\definecolor{darkblue}{rgb}{0.02, 0.16, 0.49}
\newcommand{\framework}{{Papora}}

\def\BibTeX{{\rm B\kern-.05em{\sc i\kern-.025em b}\kern-.08em
    T\kern-.1667em\lower.7ex\hbox{E}\kern-.125emX}}
\begin{document}
%
\title{Fuzzing the Latest NTFS in Linux with Papora: An Empirical Study\thanks{This work is partly supported by the National Key Research and Development Program (No. 2022YFB4501802), the National Natural Science Foundation of China (No. 62172009, No. 62141208), and Amber Group.}}


\author{\IEEEauthorblockN{Edward Lo\IEEEauthorrefmark{1}\IEEEauthorrefmark{3},
Ningyu He\IEEEauthorrefmark{2}\IEEEauthorrefmark{3},
Yuejie Shi\IEEEauthorrefmark{1},
Jiajia Xu\IEEEauthorrefmark{1},
Chiachih Wu\IEEEauthorrefmark{1},
Ding Li\IEEEauthorrefmark{2}, and
Yao Guo\IEEEauthorrefmark{2}\Letter}\\
\IEEEauthorblockA{\IEEEauthorrefmark{1}\textit{Amber Group}}
\IEEEauthorblockA{\IEEEauthorrefmark{2}\textit{Key Lab on HCST (MOE), School of Computer Science, Peking University}}
\IEEEauthorblockA{\IEEEauthorrefmark{3}Co-first authors}
}

\maketitle

\begin{abstract}
Recently, the first feature-rich NTFS implementation, NTFS3, has been upstreamed to Linux.
Although ensuring the security of NTFS3 is essential for the future of Linux, it remains unclear, however, whether the most recent version of NTFS for Linux contains 0-day vulnerabilities.
To this end, we implemented {\framework}, the first effective fuzzer for NTFS3.
We have identified and reported 3 CVE-assigned 0-day vulnerabilities and 9 severe bugs in NTFS3.
Furthermore, we have investigated the underlying causes as well as types of these vulnerabilities and bugs. We have conducted an empirical study on the identified bugs while the results of our study have offered practical insights regarding the security of NTFS3 in Linux.
\end{abstract}



%
\IEEEpeerreviewmaketitle

\section{Introduction}
\label{sec:intro}
NTFS~\cite{ntfs} was developed by Microsoft as the native file system for Windows NT.
Decades later, along with the rapid growth of the market share of Windows, numerous hard disks are formatted as NTFS, whose fully read-write support should be taken into consideration for other operating systems, e.g., Linux kernel.
NTFS3, as the first feature-rich implementation of the impactful NTFS file system, landed in Linux in late 2021.
Albeit the potential benefit, integrating a new component, especially a file system, is extremely likely to introduce bugs or even vulnerabilities.
Unfortunately, to the best of our knowledge, there is no systematic study on the found bugs introduced by the latest NTFS3.
Even worse, we find there are even no available tools to discover these bugs.
Thus, regarding integrated NTFS3 in the Linux kernel, it is necessary to implement a tool to detect bugs, and conduct a systematic evaluation on them to raise the awareness of the community, especially security researchers.

To close this gap, in this paper, we build the first effective fuzzer, named {\framework}, for NTFS3. Then, we conduct the first fuzzing-based systematic study on identified bugs.
In this whole process, we have to underline that it is particularly challenging in engineering.
Although there are several fuzzers for file systems, such as Janus~\cite{janus} and Hydra~\cite{hydra}, they cannot be directly applied to fuzz NTFS3 due to two issues.
First, they lack a specific parser for NTFS images to extract metadata, correct checksums and assemble corpus.
Moreover, directly adopting existing parsers for NTFS is not feasible because they only validate the integrity of the given image, which is insufficient for fuzzy testing.
Lacking such a parser will significantly decrease the performance of a fuzzer due to the burden on I/O issues~\cite{afl,fuzzbsd,schumilo2017kafl}.
Second, they do not support KASAN~\cite{kasan} on our targeting Linux kernel, making the vulnerability hunting less effective.

Fortunately, {\framework} has addressed the above two tough nuts to some extent.
On the one hand, since the implementation of NTFS is not open-sourced, it is particularly tough to build a feasible image parser for it.
Though there is a so-called official documentation, it still lacks lots of technical details.
To this end, we manually compare multiple third-party releases and their corresponding documentations, and cross-reference which implementation is consistent with the expected behavior.
On the other hand, because fuzzing an image via virtual instances may suffer bug reproduction issues~\cite{janus}, we decided to apply LKL~\cite{lkl}, a user space application that can emulate behaviors of the Linux kernel, to load NTFS images.
However, LKL is not maintained at all, and KASAN is not integrated inside.
Therefore, we firstly ported LKL to the latest version, and revisited the instrumented memory subsystem of LKL and enabled KASAN support with intensive effort.
For example, LKL adopts a special architecture, i.e., running with a linear memory layout, which is different from other \textit{memory management unit based} (MMU-based) architectures having KASAN support.
We have to manually refactor the KASAN codebase and make it compatible with the no-MMU LKL.

The results have proven that {\framework} is an effective fuzzer.
In total, we have discovered 3 CVE-assigned 0-day vulnerabilities and 9 severe bugs in the latest Linux kernel.
We have reported these 12 vulnerabilities/bugs to the maintainers with patches, which have all been confirmed. Moreover, 9 out of them have been merged into the upstream. 
Our study shows that the latest version of NTFS in Linux still suffers from the problems of out-of-bounds read bugs and null pointer dereference bugs.
Moreover, to ring the alarm for the community of the security issues resulting from enabling NTFS support, we have conducted sophisticated and meaningful case studies on representative identified bugs. Based on the case study, we also propose some best practices for security researchers and Linux developers to avoid such bugs.
We urge the developers to have a deeper investigation into these two types of bugs and improve their security awareness with our empirical study.

We summarize our contribution as follows:
\begin{enumerate}
	\item To the best of our knowledge, we have implemented the first fuzzer specifically for NTFS3 in Linux. It is able to effectively and efficiently discover new bugs and vulnerabilities.
	\item We have identified 3 CVE-assigned 0-day vulnerabilities and 9 severe bugs in NTFS3, among which 9 were confirmed and fixed on the upstream.
    \item We have made an empirical fuzzing-based case study on bugs of NTFS3 in Linux. 
    \item We have released {\framework} including the LKL, which is ported to the latest version of the Linux kernel and integrated with KASAN, at \href{https://github.com/ambergroup-labs/papora}{https://github.com/ambergroup-labs/papora}.
\end{enumerate}

This paper is organized as follows:
\S\ref{sec:background} introduces some basic knowledge of NTFS and fuzzing testing. In \S\ref{sec:experimental-setup}, we detail the challenges for fuzzing an NTFS image, and how we build {\framework}. Moreover, in \S\ref{sec:results}, we illustrate the 12 bugs/vulnerabilities we identified, and conduct case studies on representative ones.
Based on the results, we have summarized some best practices for users, developers and security researchers in \S\ref{sec:lessons}.
Finally, \S\ref{sec:related} and \S\ref{sec:discussion} illustrate published related work and a discussion of some interesting issues of this paper, respectively.

\section{Background}
\label{sec:background}

\subsection{NTFS File System}
A file system is one of the essential components of an operating system that 
manages files, folders, links, and the data to efficiently process the 
read/write requests from users to those items.
In 1993, the proprietary New Technology File System (NTFS)~\cite{ntfs}
developed by Microsoft debuted with the first release of Windows NT.
Within the last two decades, various NTFS drivers including the legacy driver~\cite{ntfs-rst} 
and FUSE-backed drivers, 
Captive~\cite{captive} 
and
NTFS-3G~\cite{ntfs-3g}, 
had been contributed to the Linux kernel for providing alternative ways to 
access Windows hard drivers on Linux.
In late 2021, the presence of NTFS3~\cite{ntfs3}
developed by Paragon Software~\cite{paragon} 
finally unleashes the power of NTFS in Linux 5.15 kernel.

\subsubsection{NTFS Features}
\label{sec:background:features}
As the successor of the File Allocation Table (FAT) file system, except for supporting some advanced features like large volumes, NTFS outperforms FAT in many aspects, especially the \textit{reliability} and \textit{security}.
Specifically, the reliability of NTFS file systems can be reflected from two aspects.
On the one hand, similar to other journaling file systems, NTFS uses the logging and checkpoint mechanisms to guarantee the consistency of the file system for dealing with unexpected system crashes.
On the other hand, a recovery technique, named \textit{cluster remapping}, can also improve the reliability.
When a bad sector, located in a cluster, is detected in a read operation, NTFS remaps the cluster to a newly allocated one, and marks the bad one that will no longer be used.
As for the security issue, NTFS allows granting access to files or directories in users and groups granularity.
Moreover, the Encrypting File System (EFS) allows users to encrypt files on NTFS volumes.
In this way, even a bad actor can physically access the hard drive, he cannot decrypt any files in an EFS-enabled NTFS volume without the owner's private key~\cite{efs}.

\begin{figure}[t]
    \centering
    \includegraphics[width=0.45\textwidth]{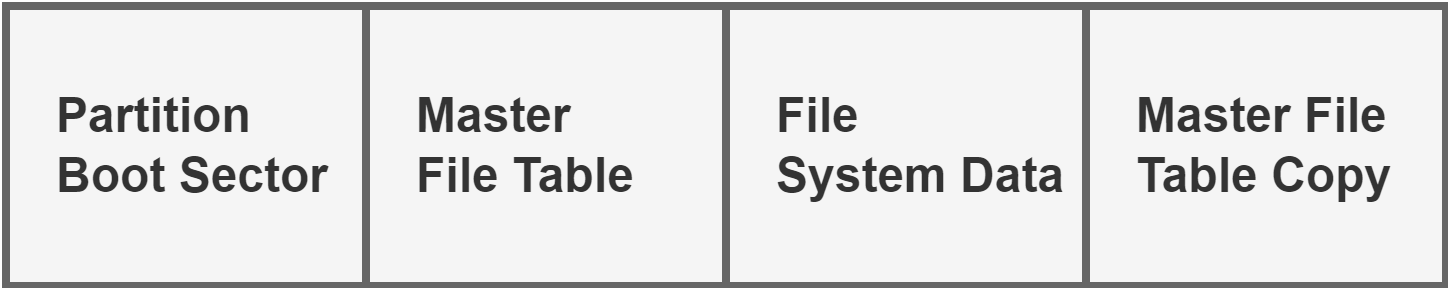}
    \caption{The Layout of an NTFS File System.}
    \label{fig:ntfs}
\end{figure}

\subsubsection{NTFS Physical Structure}
\label{sec:background:phy}
Fig.~\ref{fig:ntfs} illustrates the layout of an NTFS image.
The \textbf{Partition Boot Sector} (PBS) holds important information for 
bootstrapping the system.
In particular, the boot sector starts with an x86 \textit{jump} instruction which skips some non-executable metadata, e.g., \textit{OEM ID}.
And the PBS is also responsible for modifying the program counter to the \textit{Bootstrap Code}.
\textbf{Master File Table} (MFT) holds the metadata of all files and directories, even including the metadata itself.
To ensure the integrity of MFT, NTFS maintains a \textbf{Master File Table Copy}, which maintains exact identical data to the MFT.
The MFT is composed of multiple entries for NTFS metadata, each of which has fixed functionalities and follows strict syntax.
For example, the third entry of the MFT, named \textit{\$LogFile}, contains all necessary transactions for a faster recovery when the system crashes. Moreover, the seventh entry, dubbed as \textit{\$Bitmap}, maintains a bitmap for all free and unused clusters.
Most importantly, MFT also stores some meta-information necessary to retrieve files, like the attributes of a file.
In NTFS, each file is stored in clusters that are composed of one or multiple sectors, and structured in a list of attributes, e.g., file name, timestamp, even the file data.
The file data that is not contained in the MFT will be stored in \textbf{File System Data}~\cite{ntfs-org}.

\subsection{Fuzzing}
\label{sec:background:fuzzing}
\textit{Fuzzing} or \textit{fuzzy testing} is an automatic software testing paradigm that tests the 
target with inputs which are mutated based on the target state and testing 
results.
A naive practice is randomly generating inputs to fuzz the target until it 
crashes.
It might work when the input space is limited but many popular fuzzers guide the 
input mutation based on code coverage.
Coverage-based fuzzers such as AFL~\cite{afl} and libFuzzer~\cite{libfuzzer}  
instrument the target in compile 
time and feed the target states back to the input mutator, which leads the 
fuzzer to keep exploring new execution paths in the target program.

While fuzzing a user space program, the target takes the mutated inputs from the 
command line or configuration files and executes in a loop until it crashes.
But, it is a different story to fuzz an operating system component like a 
file system.
Specifically, the input space becomes two dimensions, i.e., an image with a file system, and a series of system calls.
Traditional kernel fuzzers such as Trinity~\cite{trinity} and Syzkaller~\cite{syzkaller} 
generate a series of system calls with parameters as inputs for the target 
operating system.
In particular, Trinity uses the annotation to generate \textit{better-than-random}
parameters for each system call to trigger unexpected behaviors more easily.
Syzkaller uses KCOV to collect code coverage and SyzLang to 
provide context for guided fuzzing.
However, the file system image is a more complicated input that cannot be 
efficiently generated by kernel fuzzers.

File system fuzzing efforts such as Janus~\cite{xu2019fuzzing} and 
Hydra~\cite{kim2020finding} deal with the problem by extracting metadata from 
large images with file system-specific parsers. 
Janus also addresses the aging OS problem with the library OS LKL~\cite{lkl}, 
which helps the fuzzer to quickly reload a clean-slate OS and get rid of 
irreproducible bugs.

\section{Experimental Setup}
\label{sec:experimental-setup}
In this section, we will explain how we implement an NTFS fuzzer and how to efficiently fuzz an NTFS file image.

\subsection{Challenges}
\label{sec:experimental-setup:challenges}
As we mentioned in \S\ref{sec:background:fuzzing}, fuzzing a file system is challenging, which can be summarized as follows: 

\begin{itemize}
	\item \textbf{C1: Disk Image.} For a mainstream fuzzer, e.g., AFL~\cite{afl}, its recommended size of targets is less than 1KB. An empty file system, however, which is embedded in a disk image, often contains more than dozens of megabytes. Directly fuzzing a disk image, which is 1,000x larger than fuzzer's maximum preferred size, will dramatically downgrade the efficiency due to the heavy I/O brought by mutating or booting the given image.
	\item \textbf{C2: Context-aware File Operations.} Except for directly mutating the disk image, file operation is another orthogonal valuable seed. In other words, a series of file operations may also lead to system crashes. Moreover, file operations are context-aware workloads for images. For example, \texttt{open()} will actually create a new file on the target image, based on which the following file operations can be performed. Such context-aware file operations not only exponentially increase the exploration space for seed mutation, but also require the corresponding updates on states (e.g., entries in MFT) of the image.
	\item \textbf{C3: Reproduction.} Traditional fuzzers aiming at operating systems often take virtual instances as targets. However, frequently modifying and rebooting virtual instances, or reverting to specific snapshots are extremely time-consuming. Moreover,  they may reuse file systems, leading to undetermined and unpredictable states, i.e., aging problems, for file systems, which seriously hinders the reproduction of found bugs.
\end{itemize}

Janus and Hydra have addressed the above challenges to some extent on multiple file systems, e.g., ext4 and HFS+, except for NTFS.
Compared with those targets, the biggest obstacle of fuzzing an NTFS image is the absence of its implementation.
Except for Microsoft's documentation where it qualitatively describes the structure and implementation of NTFS, all other releases as we mentioned in \S\ref{sec:intro} are third-party implementations.
To this end, efficiently and correctly fuzzing an NTFS image is challenging.

\subsection{Overview}
\label{sec:experimental-setup:overview}
\begin{figure*}[t]
    \centering
    \includegraphics[width=0.9\textwidth]{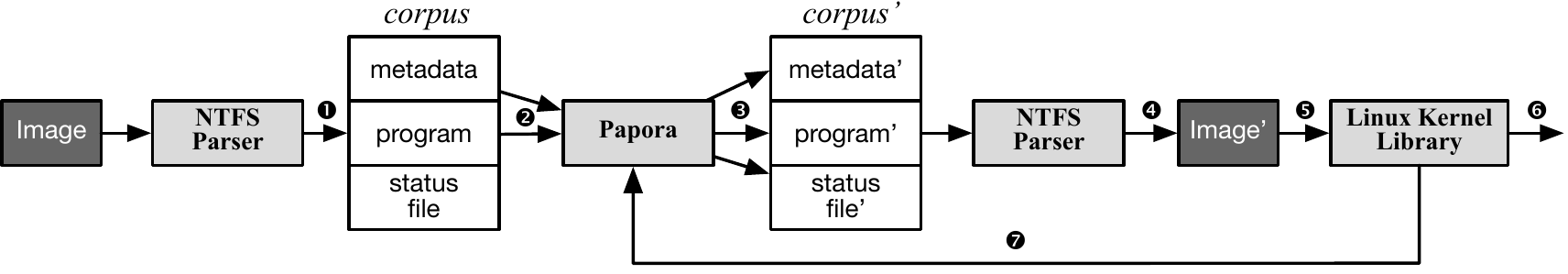}
    \caption{Overall workflow of {\framework}.}
    \label{fig:overview}
\end{figure*}

According to challenges we introduced in \S\ref{sec:experimental-setup:challenges}, {\framework} is specifically designed to tackle these problems. The overall workflow of {\framework} is shown in Fig.~\ref{fig:overview}.

As we can see, firstly, an NTFS parser scans the whole given image, and builds a corpus that will be sent to {\framework}. Our fuzzer tries to mutate both the metadata of the given image and the program consisting of file operations, and updates the status field accordingly (Step 2 \& 3).
Then, the NTFS parser assembles the updated corpus as an intact mutated image (Step 4), which will be mounted by Linux Kernel Library (LKL) and executed according to the program (Step 5).
Finally, the corresponding result of the current round will be outputted (Step 6), and the feedback information will be sent back to {\framework} to guide the following mutations (Step 7).
Technical details are illustrated from the following \S\ref{sec:experimental-setup:preparation} to \S\ref{sec:expremental:lkl}.

\subsection{Corpus Building}
\label{sec:experimental-setup:preparation}
As we can see from \S\ref{sec:experimental-setup:overview}, a corpus is composed of three parts: extracted metadata of the given image, a program consisting of a series of file operations, and a status file.
Specifically, a specifically-designed \textit{NTFS parser} (see \S\ref{sec:experimental-setup:parser}) will extract the metadata, i.e., entries in PBS and MFT, out of the image, and condense them into a bulk of data. In this way, the meaningless part for finding new bugs, i.e., File System Data (see Fig.~\ref{fig:ntfs}), contributing more than 99\% of space to the image, will not be included into the corpus.
As for the second part, the initial program is an empty sequence of file operations.
Last, as we mentioned in \S\ref{sec:background:phy}, attributes of files and directories are stored in the MFT. During scanning the image, these attributes will be packed and maintained in the third part, i.e., the status file, of the corpus.
{\framework} will take the assembled corpus as input, mutate either the metadata or the file operations, mount the image, and execute the program to see if any bugs are triggered. If it is not, the corresponding field of the input corpus will be updated (like the status field should be updated due to file creations), and the corpus will be sent to {\framework} for the next round fuzzing.

\subsection{NTFS Parser}
\label{sec:experimental-setup:parser}
As we mentioned in \textbf{C1}, directly fuzzing an image will face an extreme efficiency problem.
Therefore, like previous work~\cite{janus,hydra}, we develop a \textit{parser} specifically targeting NTFS to tackle this problem.
The responsibility of the parser can be divided into three-folds.

\textit{First}, the parser can extract all metadata and compress them into a dense bulk of data. For a file system, an image crash after mounting is only due to buggy metadata, accounting for less than 1\% space of the image.
This indicates that mutating the other 99\% space (mainly composed of files' content) is meaningless.
As for an NTFS image, metadata is mainly composed of fields in \textit{Partition Boot Sector (PBS)} and \textit{Master File Table (MFT)} (see \S\ref{sec:background:phy}).
Therefore, condensing metadata in PBS and MFT where the mutation plays on will not only increase the efficiency for both mutation and the following fuzzing, but also increase the possibility of finding corrupted metadata related new bugs.

\textit{Second}, the parser will automatically fix checksums after mutating metadata. For file systems, including NTFS, they all adopt checksums to ensure the integrity and usability of the metadata. A mismatch between checksum literals and in-time calculated checksums will raise an error, resulting in the image cannot be loaded properly. Therefore, after mutating metadata, the parser will re-calculate all the corresponding checksums to guarantee the mutated NTFS file system can pass the static verification on checksums.
From \S\ref{sec:background:phy}, we can see that the PBS and MFT are NTFS-specific structures. {\framework} has some specific checksum fixups on these two data structures.
Specifically, the PBS holds important information for boot (see \S\ref{sec:background:phy}). For example, its second field, i.e., OEM ID, is fixed as ``NTFS'' followed by 4 space characters. Though this field is mutated by {\framework} accidentally, the parser will recover its original value to make sure mutation has no negative effects on booting.
Additionally, MFT stores metadata of all files and directories. The header of each file record contains a \textit{Update Sequence Number} (USN) and a buffer. NTFS requires the last two bytes of each sector of records are copied into the buffer and the USN is written in their place. After booting, NTFS will compare the USN from the header with the last two bytes of each sector. To this end, if {\framework} mutates headers in MFT, it will modify the corresponding fields to pass sanity checks.

\textit{Third}, the fuzzing process still performs on an intact image, thus the parser should also be responsible for mapping the mutated metadata in corpus back to the image.
To achieve such a goal, during the extracting, the parser will maintain a bitmap in which the offsets of each piece of metadata are kept. Based on the bitmap, the mutated metadata can be filled back into the original slots.

Note that, instead of directly adopting traditional and well-maintained NTFS parsers, like NTFS-3G, implementing our own parser does not mean reinventing a wheel.
Although traditional parsers can parse and load NTFS images, they only validate if the given image is valid, or the image will not be loaded successfully.
However, as we mentioned above, our parser is responsible for correcting checksums and assembling corpus for the following process. In other words, our parser conducts an extra fixup process based on verifying the validity.
Moreover, NTFS-3G is too heavy and inefficient for fuzzing analysis.
In summary, it is necessary to implement our own parser in implementing {\framework}.

\subsection{Fuzzing Image}
{\framework} applies several strategies, e.g., bit/byte flip, and arithmetic operations, on the metadata part of a corpus to mutate it. The strategies can be summarized as follows:

\begin{itemize}
    \item Flip a bit at random offset, or set an interesting byte/word/dword value (like min/max valid number, 0, $\pm$1, and power of 2) at random offset in random endian.
	\item Randomly add/subtract a random value at random byte/word/dword offset.
    \item Overwrite bytes by a random chunk or a random byte for random length at random offset, or by user specified tokens if provided.
\end{itemize}

After mutating the metadata, the NTFS parser recalculates necessary checksums.

\subsection{Fuzzing File Operations}
Except for mutating metadata, {\framework} also mutates the second part of corpus, i.e., the program consisting of a series of file operations. {\framework} mutates the program in two strategies: \textit{mutation} and \textit{generation}.

\noindent
\textbf{Mutation.} {\framework} prefers this strategy. It randomly picks one file operation in the seed program, then replaces some of its arguments with heuristic values instead of random ones.
As we stated in \textbf{C2}, these file operations are context-aware. Thus, the selected values should be meaningful for the current image.
For example, if the mutated file operation is \texttt{fsync()}, which takes a file descriptor as its argument to synchronize its in-core state with the storage device. {\framework} will pick one of the \textit{opened} file descriptors of proper type.
Such a heuristic and context-aware strategy also applies for mutating system calls related to path and extend attributes.

\noindent
\textbf{Generation.} If {\framework} cannot increase coverage through mutating the program, it will try to append new file operations to the program with proper arguments.
Moreover, the potential side effects of each file operation are taken into consideration, and the program context will be updated accordingly.
For instance, \texttt{link()} and \texttt{mkdir()} may create a new file and directory, while \texttt{unlink()} and \texttt{rmdir()} have the opposite effects. The program context will record changes introduced by these system calls in the status file of corpus.

\subsection{Linux Kernel Library (LKL)}
\label{sec:expremental:lkl}
\textbf{C3} has stated that if {\framework} adopts virtual instances to mount the image and execute the program, it will face efficiency and reproduction problems.
To this end, {\framework} builds its target program, i.e., the executor, with \textit{Linux Kernel Library} (LKL)~\cite{lkl}, as a user space program.
LKL provides a way for emulating the Linux kernel by compiling the kernel into an object file that can be directly linked by applications.
To discover potential vulnerabilities in the latest version, we have upgraded LKL to v6.0\footnote{At the time of writing, v6.0 is the latest version for Linux. Moreover, the LKL project was inactive and the supported kernel version stayed at v5.3.}.
Moreover, to enable detection of illegal memory accesses, we also integrated the \textit{KASAN}~\cite{kasan} provided by~\cite{hydra} into the LKL with several necessary fixups.

Porting the LKL to the latest Linux kernel and integrating KASAN into it are challenging.
On the one hand, some interfaces have been introduced, changed or even deprecated. For example, the interface of \texttt{copy\_thread} is changed, leading to a rewriting of the corresponding function to guarantee the logic correctness. Moreover, header files rearrangement often happens in the upstream. Thus, subsystem maintainers may choose to move some structures or macro definitions to new headers, which may lead to merge conflicts or build errors while porting LKL to the latest kernel.
On the other hand, integrating KASAN into LKL needs lots of effort. This is because LKL can be regarded as an architecture with no-MMU (memory management unit) support. In other words, LKL only supports linear memory address, which is conflicted with KASAN initialization flow. Therefore, we have to manually review the flow, adjust or comment out related codes or structures, while maintaining the functionalities of KASAN.

After resolving problems of porting LKL to the latest version, it can bring in several advantages over mounting images through a virtual machine.
First, user-space applications are much lighter than the emulator in terms of rebooting. Restarting an application only introduces negligible time compared to resetting a VM instance.
Second, VM-based fuzzers may choose to keep running their target programs until the \textit{aging} kernel crashes or hangs. To this end, the initial status of the image is undetermined, which results in irreproducible bugs even with full kernel dump. For security researchers, it is also difficult to obtain the root cause in such indeterministic situations. {\framework}, however, can restart its executor for every corpus with little overhead, providing a stable and determined kernel state.
Third, such a LKL assisted method requires much less computing resources, so it is easy to scale up the fuzzing process by deploying more instances.

\section{Experiment Results}
\label{sec:results}
In this section, we will first list all bugs identified by {\framework}. Then, we will delve deeper and conduct case studies to illustrate the reason behind system crashes.
The results show that some severe vulnerabilities may even be used in privilege escalation.

\subsection{Results}

\begin{table}[t]
\centering
\caption{Identified bugs and vulnerabilities (highlighted rows) by {\framework}, where NPD and OOB refer to null pointer dereferences and out-of-bound, respectively.}
\resizebox{\columnwidth}{!}{%
\begin{tabular}{cccc}
    \toprule
    Commit & Bug Type & Root Cause & Upstreamed\\
    \midrule
    \multicolumn{4}{l}{\textbf{Type I}} \\
    0b66046 & NPD & Sanity check miss & \Checkmark\\
    e19c627 & OOB Read & Arithmetic overflow & \Checkmark\\
    6db6208 & OOB Read & Sanity check miss & \Checkmark\\
    2681631 & NPD & Sanity check miss & \Checkmark\\
    c1ca8ef & NPD & Implementation flaw & \Checkmark\\ 
    \rowcolor[HTML]{EFEFEF}
    \begin{tabular}[c]{@{}c@{}}4f1dc7d\\ (\textbf{CVE-2022-48424})\end{tabular} & Heap Corruption & Sanity check miss & \Checkmark\\
    bfcdbae & OOB Read & Sanity check miss & \Checkmark\\
    e6ffad3 & OOB Read & Sanity check miss &\\
    \rowcolor[HTML]{EFEFEF}
    \begin{tabular}[c]{@{}c@{}}467333a\\ (\textbf{CVE-2022-48425})\end{tabular} & Heap Corruption & Type confusion &\\
    f64633f & OOB Read & Sanity check miss &\\
    \midrule
    \multicolumn{4}{l}{\textbf{Type II}} \\
    4d42ecd & OOB Read & Sanity check miss & \Checkmark\\
    \rowcolor[HTML]{EFEFEF}
    \begin{tabular}[c]{@{}c@{}}54e4570\\ (\textbf{CVE-2022-48423})\end{tabular} & OOB Write & Sanity check miss & \Checkmark\\
    \bottomrule
\end{tabular}%
}
\label{table:Trophy} 
\end{table}

We run {\framework} on a VMware virtual machine with an 8-core CPU and 16GB
memory running Ubuntu 16.04.
The experiment is conducted intermittently for about 3 months.
As the results listed in Table~\ref{table:Trophy}, {\framework} has successfully discovered 9 severe bugs and 3 CVE-assigned vulnerabilities (highlighted rows) in the NTFS3 implementation\footnote{For simplicity, all these 12 identified issues will be referred by \textit{bugs} if they are mentioned as a whole.}.
All identified bugs as well as the corresponding patches have been reported to maintainers, and 9 of them have been merged into the upstream.
Additionally, we have categorized these identified bugs into two types.
The \textit{Type I} refers to the situation that once the NTFS image is mounted, the system crashes.
While the ones under \textit{Type II} can only be triggered by invoking the corresponding system calls after mounting the image.

\subsubsection{Categorized by Bug Type}
Over 60\% of the bugs identified by {\framework} are out-of-bounds read bugs.
Those bugs are the most dangerous species which could be exploited for leaking 
kernel information or even corrupting kernel memory.
For example, an out-of-bounds read could be used to export the addresses of critical kernel 
data structures to be corrupted.
By exploiting an out-of-bounds write, a process with the mounting capability 
could escalate its privileges by corrupting function pointers, hijacking the 
control flow (e.g., jumping to the shellcode or JOP gadgets), and eventually 
changing credentials data in the \texttt{task\_struct}.

Around 25\% of the bugs identified by {\framework} are null pointer dereference (NPD) bugs.
Those bugs directly crash the target system and make the target hang or reboot 
depending on the system configuration.
By exploiting an NPD, a bad actor could launch denial-of-services attacks on 
target systems with the mounting capability or the auto-mounting feature enabled.

Except for OOB access and NPD bugs, {\framework} identified 2 heap corruption 
bugs which could be developed into use-after-free (UAF) exploits.
Specifically, when a memory chunk is allocated in the Linux kernel, a reference pointer is 
returned by the slab system and all upcoming access would go through that pointer.
A typical exploit is filling another victim memory chunk containing function 
pointers into the intentionally released spot and corrupting the victim memory 
chunk through the old reference pointer.

\subsubsection{Categorized by Root Cause}
\label{sec:expr_results:cause}
While analyzing those bugs identified by {\framework}, most of them are due to 
missed sanity checks on user-controllable data.
Specifically, any data field retrieved from a file system image is a chunk of 
user-controllable data which should always be strictly checked as it would be 
used as the input of the NTFS3 implementation.
For example, a crafted \texttt{offset} field could simply lead to an 
out-of-bounds read if it is not bounded by the size of the allocated memory to 
cache the metadata.
Furthermore, if the \texttt{offset} is derived from another 
number-of-entries field, an overflowed \texttt{offset} could be crafted 
when $\textit{number-of-entries} \times \textit{size-of-entry}$ is large 
enough.
That overflowed \texttt{offset} would bypass the sanity check for the bounding 
\texttt{offset} itself.
As a result, all arithmetic operations related to user-controllable data 
should be carefully validated as well.

{\framework} also identified bugs caused by type confusion~\cite{cwe-843}. 
We believe that it is a common type of bug in Linux file system 
implementations due to the design of inode.
In particular, each inode could be interpreted in various ways depending on the states or flags.
As shown in Listing~\ref{lst:type-confusion}, the \texttt{union} in 
\texttt{struct ntfs\_inode} makes each \texttt{ntfs\_inode} represent either a 
\texttt{dir} or a \texttt{file}.
Commit \texttt{467333a}~\cite{commit-467333a}
demonstrates a type confusion case in which the NTFS3 
implementation wrongly interprets an \texttt{MFT\_REC\_MFT} file as a 
directory and corrupts the heap by \texttt{kfree()}-ing an invalid pointer.

\begin{lstlisting}[language=C, caption={A code snippet of struct \texttt{ntfs\_inode}}, label={lst:type-confusion}]
union {
  struct ntfs_index dir;
  struct {
    struct rw_semaphore run_lock;
    struct runs_tree run;
#ifdef CONFIG_NTFS3_LZX_XPRESS
    struct page *offs_page;
#endif
  } file;
};
\end{lstlisting}

The root cause of \texttt{c1ca8ef} bug identified by {\framework} could be categorized in the \textit{Always-Incorrect Control Flow Implementation} class~\cite{cwe-670}.
In other words, instead of preparing a malicious input, a bad actor could trigger the crash with 
a normal test case which is missed due to incomplete test coverage.
That is exactly the problem we need a customized fuzzer like {\framework} to cope with.

\subsection{Case Study on Type I}
\label{sec:result-typei}
\begin{figure}[t]
    \centering
    \includegraphics[width=0.45\textwidth]{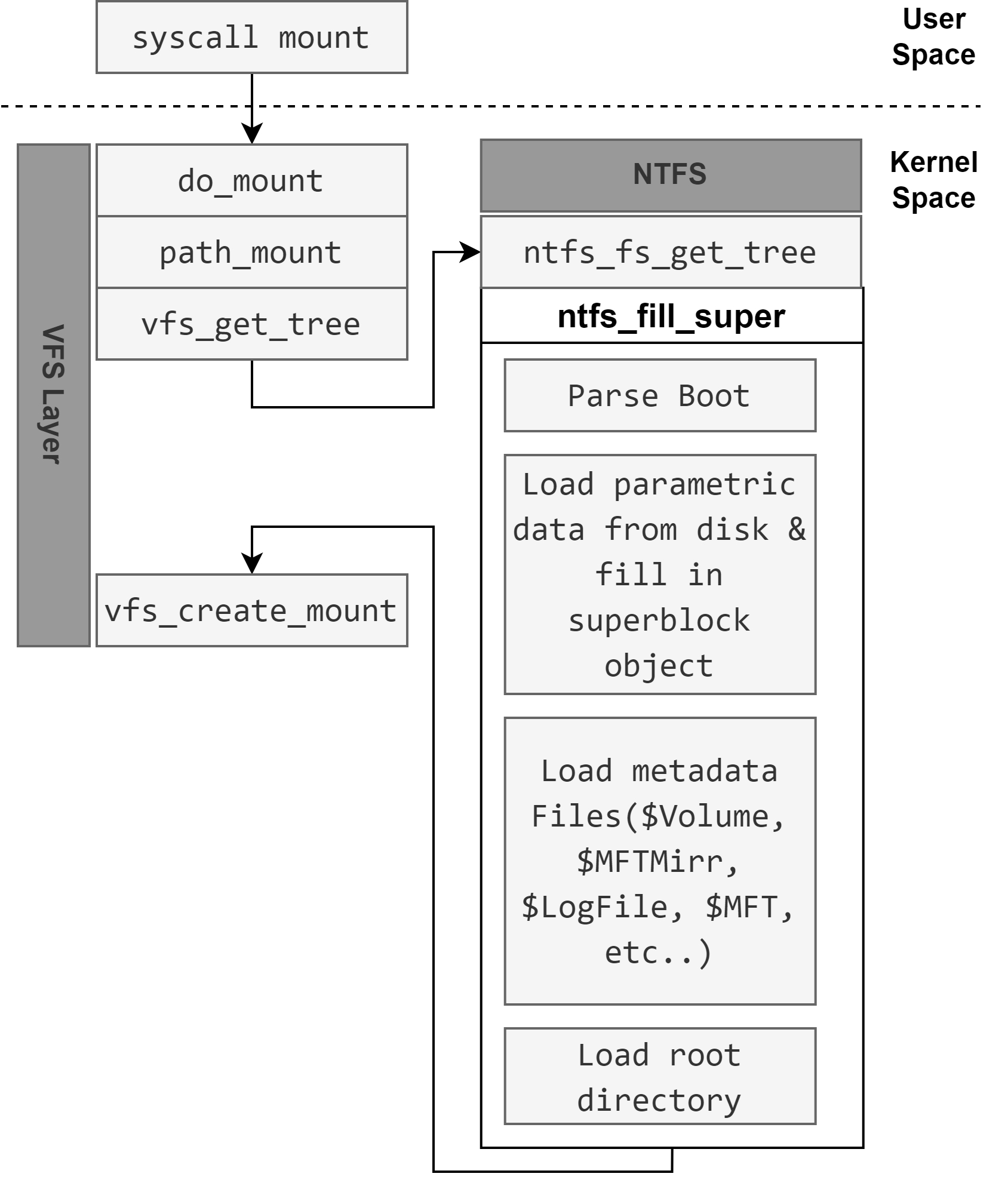}
    \caption{The simplified trace of mounting an NTFS disk.}
    \label{fig:mount}
\end{figure}

Type I bugs occur during mounting an NTFS disk, whose traces are shown in Fig.~\ref{fig:mount}.
As we can see, once invoking the \texttt{mount}, the Linux system will trap into the kernel space.
Most of the mounting processes are handled by Linux's VFS layer.
Because files in Linux are arranged in a tree-like hierarchical structure, the \texttt{vfs\_get\_tree} will call the specialized \texttt{ntfs\_fs\_get\_tree} to get its mountable root.
Within the implementation of NTFS, the function \texttt{ntfs\_fill\_super} plays a vital role.
Specifically, it parses the partition boot sector (see \S\ref{sec:background:phy}) and reads parametric data, e.g., cluster size and maximum size of normal files.
It also loads all metadata files from the master file table. Finally, it reads the root directory of the NTFS file system from disk.
All these loaded data will be filled into a superblock structure, i.e., \texttt{ntfs\_sb\_info}.

In this section, we conduct case studies against three representative Type I bugs, i.e., \texttt{0b6604}, \texttt{c1ca8e}, and \texttt{e19c62}.
The root causes for these three bugs are different. However, the system will crash once the image is mounted.
We will delve deeper in the following.
 
\subsubsection{\texttt{0b66046}}
\label{sec:typei-1}
This bug results from a null pointer dereference due to an implementation error.
Specifically, as we mentioned in Fig.~\ref{sec:result-typei}, the first step of \texttt{ntfs\_fill\_super} is parsing the partition parse boot, which is implemented through a function named \texttt{ntfs\_init\_from\_boot}, which is implemented in Listing~\ref{lst:ntfs-init-from-boot}.

\begin{lstlisting}[language=C, caption={The implementation of \texttt{ntfs\_init\_from\_boot}}, label={lst:ntfs-init-from-boot}]
static int ntfs_init_from_boot(struct super_block * sb, u32 sector_size, u64 dev_size) {
  // some operations
  sbi -> record_size = record_size = boot -> record_size < 0 ?
  1 << (-boot -> record_size) :
  (u32) boot -> record_size << sbi -> cluster_bits;

  if (record_size > MAXIMUM_BYTES_PER_MFT)
    goto out;

  sbi -> record_bits = blksize_bits(record_size);
  // some operations
}

/* assumes size > 256 */
static inline unsigned int blksize_bits(unsigned int size) {
  unsigned int bits = 8;
  do {
    bits++;
    size >>= 1;
  } while (size > 256);
  return bits;
}
\end{lstlisting}

As we can see, the boot record size (\texttt{boot->record\_size}) is read from disk at L3, which will then be passed to function \texttt{blksize\_bits} through a variable named \texttt{record\_size}.
The comment at L14 says that \texttt{record\_size} should be larger than 256.
However, before passing \texttt{record\_size} to \texttt{blksize\_bits}, there is only a verification to identify if it is greater than a maximum limit (L7).
In other words, if \texttt{record\_size} is less than 256, function \texttt{blksize\_bits} will only return 8 to \texttt{record\_size} (L10), smaller than the ordinary situations.
In the following stages during mounting the disk, a pointer will be shifted left \texttt{record\_bits} bits and shifted right several bits.
Because the variable \texttt{record\_bits} will be only 8 when \texttt{record\_size} is smaller than 256, the value of the pointer will be so small that it will point to a invalid address, leading to a null pointer dereference.
According to the log file, the value of the pointer, i.e., the address, is only \texttt{0000000000000158}, which is an invalid memory address.

Therefore, the patch will limit the range of \texttt{record\_size}.
In other words, its acceptable range should be limited by not only a maximum value (\texttt{MAXIMUM\_BYTES\_PER\_MFT} at L7 in Listing~\ref{lst:ntfs-init-from-boot}), but also by a minimum value.
Listing~\ref{lst:patch-typei-1} shows the corresponding patch.
As we can see, we set a minimum limit as \texttt{SECTOR\_SIZE}, because a boot record in NTFS always includes the first sector of the disk image, whose size is \texttt{SECTOR\_SIZE}, i.e., 512 bytes.

\begin{lstlisting}[language=C, caption={Patch to the bug in \S\ref{sec:typei-1}}, label={lst:patch-typei-1}]
diff --git a/fs/ntfs3/super.c b/fs/ntfs3/super.c
index d72a27abf1c83..af9b7947df64e 100644
--- a/fs/ntfs3/super.c
+++ b/fs/ntfs3/super.c
@@ -814,7 +814,7 @@ static int ntfs_init_from_boot(struct super_block *sb, u32 sector_size,
                          : (u32)boot->record_size
                                << sbi->cluster_bits;
 
-    if (record_size > MAXIMUM_BYTES_PER_MFT)
+    if (record_size > MAXIMUM_BYTES_PER_MFT || record_size < SECTOR_SIZE)
         goto out;
 
     sbi->record_bits = blksize_bits(record_size);
\end{lstlisting}

\subsubsection{\texttt{e19c627}}
\label{sec:typei-2}
Commit \texttt{e19c627} is related to another Type I bug.
It will eventually lead to an out-of-bound access due to a missing on integer overflow check.

Specifically, this bug lies in the function \texttt{mi\_enum\_attr}.
It is an enumerator on file attributes of the disk image.
As we mentioned in \S\ref{sec:background:phy}, the MFT data structure centrally maintains attributes of files, whose detailed implementation is upon a struct, named \texttt{ATTRIB}, which is implemented in Listing~\ref{lst:attrib}.

\begin{lstlisting}[language=C, caption={The data structure maintains attributes of files}, label={lst:attrib}]
struct ATTRIB {
  enum ATTR_TYPE type;  // 0x00: The type of this attribute.
  __le32 size;    // 0x04: The size of this attribute.
  u8 non_res;    // 0x08: Is this attribute non-resident?
  u8 name_len;    // 0x09: This attribute name length.
  __le16 name_off;  // 0x0A: Offset to the attribute name.
  __le16 flags;    // 0x0C: See ATTR_FLAG_XXX.
  __le16 id;    // 0x0E: Unique id (per record).

  union {
    struct ATTR_RESIDENT res;     // 0x10
    struct ATTR_NONRESIDENT nres; // 0x10
  };
};
\end{lstlisting}

The field named \texttt{size} (L3) records the size of this struct.
Because this struct lies on the disk one by one adjacently, through reading the field \texttt{size} of the current struct, the system is able to get the address of the next struct.
Intuitively, an illegal access tends to happen if the \texttt{size} is too big.
In \texttt{mi\_enum\_attr}, the size of a struct will be directly assigned to a variable named \texttt{asize}. To get the next struct's offset on disk, \texttt{asize} is added with the offset of the current struct, dubbed \texttt{off}.
If \texttt{size} is too big, accessing the next struct tends to fall out of the disk image. This case is considered and checked in L7 of Listing~\ref{lst:mi_enum_attr}.

\begin{lstlisting}[language=C, caption={Boundary check of \texttt{asize}}, label={lst:mi_enum_attr}]
struct ATTRIB *mi_enum_attr(struct mft_inode *mi, struct ATTRIB *attr)
{
   u32 t32, off, asize;
   asize = le32_to_cpu(attr->size);

   /* Check boundary. */
   if (off + asize > used)
       return NULL;
   ...
}
\end{lstlisting}

But if \texttt{size} is big enough, \texttt{off + asize} will overflow and generate a quite small number, leading to a failure on such a boundary check.
Thus, there will be an out-of-bound read.

The patch for this bug is straightforward. It adds another check if an integer overflow happens on the addition as shown in Listing~\ref{lst:patch-typei-2}.

\begin{lstlisting}[language=C, caption={Patch to the bug in \S\ref{sec:typei-2}}, label={lst:patch-typei-2}]
diff --git a/fs/ntfs3/record.c b/fs/ntfs3/record.c
index c8741cfa421fe..66eb11e0965ef 100644
--- a/fs/ntfs3/record.c
+++ b/fs/ntfs3/record.c
@@ -220,6 +220,11 @@ struct ATTRIB *mi_enum_attr(struct mft_inode *mi, struct ATTRIB *attr)
                return NULL;
            }
    
+        if (off + asize < off) {
+            /* overflow check */
+            return NULL;
+        }
+
            attr = Add2Ptr(attr, asize);
            off += asize;
        }
\end{lstlisting}

\subsubsection{\texttt{c1ca8ef}}
\label{sec:typei-3}
Differing from the above two cases that are logical bugs, this one is an unhandled corner case.

Specifically, this bug also happens in the function \texttt{ntfs\_fill\_super}, part of which is shown in Listing~\ref{lst:ntfs-fill-super}.
As we can see, at L1, it invokes \texttt{ntfs\_iget5} to retrieve an \texttt{inode}, which will then be dispatched into \texttt{d\_make\_root} (an API of Linux kernel's VFS subsystem) to create the root directory of the mounting disk.

\begin{lstlisting}[language=C, caption={A code snippet of \texttt{ntfs\_fill\_super}}, label={lst:ntfs-fill-super}]
    inode = ntfs_iget5(sb, &ref, &NAME_ROOT);
    if (IS_ERR(inode)) {
        ntfs_err(sb, "Failed to load root.");
        err = PTR_ERR(inode);
        goto out;
    }

    sb->s_root = d_make_root(inode);
\end{lstlisting}

The \texttt{inode} is a kernel struct that represents a file or a directory in Linux kernel.
One of the fields of this struct is named \texttt{i\_op}. It is a pointer pointing to a function table that is composed of file operation handlers like \texttt{rename}, \texttt{mkdir} and \texttt{unlink}.
When initiating an \texttt{inode}, the \texttt{i\_op} will be firstly initiated as a NULL pointer.
Then, there is a verification that can be jumped over by constructing arguments.
If it is, the control flow will be directed to a label where the \texttt{inode} will be returned directly, without assigning a concrete value for \texttt{i\_op}.
Thus, the returned \texttt{inode} will carry an empty \texttt{i\_op} and be passed to \texttt{d\_make\_root} (L8 of Listing~\ref{lst:ntfs-fill-super}), within which some file operations will be performed by dereferencing \texttt{i\_op}, the NULL pointer.
An invalid memory access is exploited.

The bug fix takes the value of \texttt{i\_op} into consideration, as illustrated in Listing~\ref{lst:patch-typei-3}.

\begin{lstlisting}[language=C, caption={Patch to the bug in \S\ref{sec:typei-3}}, label={lst:patch-typei-3}]
diff --git a/fs/ntfs3/super.c b/fs/ntfs3/super.c
index ff70e2a5f3acb..1e2c04e48f98f 100644
--- a/fs/ntfs3/super.c
+++ b/fs/ntfs3/super.c
@@ -1286,9 +1286,9 @@ load_root:
     ref.low = cpu_to_le32(MFT_REC_ROOT);
     ref.seq = cpu_to_le16(MFT_REC_ROOT);
     inode = ntfs_iget5(sb, &ref, &NAME_ROOT);
-    if (IS_ERR(inode)) {
+    if (IS_ERR(inode) || !inode->i_op) {
         ntfs_err(sb, "Failed to load root.");
-        err = PTR_ERR(inode);
+        err = IS_ERR(inode) ? PTR_ERR(inode) : -EINVAL;
         goto out;
     }
\end{lstlisting}

\subsection{Case Study on Type II}

As we mentioned in \S\ref{sec:result-typei}, when mounting an NTFS image, Linux parses some structures and loads metadata which will then be filled into an NTFS superblock.
However, the attributes of files will only be read during the corresponding file operations (like open or renaming).
Therefore, even if a disk is mounted successfully, the system may also crash when some specific file operations are invoked.
We categorize these bugs as Type II ones.
{\framework} has successfully identified two Type II bugs, which are detailed in the following.

\subsubsection{\texttt{54e4570}}
\label{sec:typeii-1}
\begin{figure}[t]
    \centering
    \includegraphics[width=0.45\textwidth]{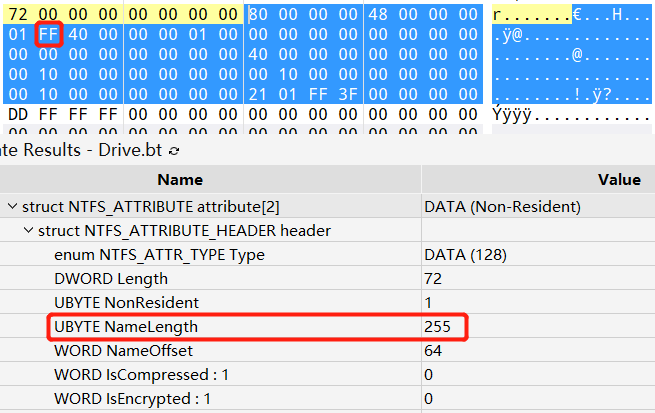}
    \caption{The mutated file attributes of the vulnerability in \S\ref{sec:typeii-1}.}
    \label{fig:hex}
\end{figure}

Commit \texttt{54e4570} is related to a Type II vulnerability\footnote{This one is not a bug because it can lead to an out-of-bound write.}, which will be triggered once updating attributes of a specific file, whose metadata is mutated by {\framework}.
Fig.~\ref{fig:hex} shows the mutated attributes of the file.
We can see that an attribute, named \texttt{NameLength}, has a value of 255.

The function \texttt{ni\_create\_attr\_list} iterates file attributes in an NTFS image with \texttt{mi\_enum\_attr} in a for-loop, as shown at L9 of Listing~\ref{lst:typeii-2}.
Then, it will copy the attributes one by one to the heap memory.
All attributes of a file are read by \texttt{mi\_enum\_attr}, but it fails to check the attribute \texttt{NameLength}. If \texttt{NameLength} is larger than the remaining allocated heap memory, a heap out-of-bound access will happen.

\begin{lstlisting}[language=C, caption={The implementation of \texttt{ni\_create\_attr\_list}}, label={lst:typeii-2}]
int ni_create_attr_list(struct ntfs_inode * ni) {
    ...
    le = kmalloc(al_aligned(rs), GFP_NOFS);
    if (!le) {
      err = -ENOMEM;
      goto out;
    }
    ...
    for (; (attr = mi_enum_attr( & ni -> mi, attr)); le = Add2Ptr(le, sz)) {
      sz = le_size(attr -> name_len);
      le -> type = attr -> type;
      le -> size = cpu_to_le16(sz);
      le -> name_len = attr -> name_len;
      le -> name_off = offsetof(struct ATTR_LIST_ENTRY, name);

      if (attr -> name_len)
        memcpy(le -> name, attr_name(attr), sizeof(short) * attr -> name_len);
    }
  }
}
\end{lstlisting}

To trigger this vulnerability, the program shown in Listing~\ref{lst:typeii-1} provides a feasible exploit.
Specifically, the program invokes \texttt{setxattr} at L16, setting attributes of a file, which eventually invokes the \texttt{ni\_create\_attr\_list} to exploit the vulnerability.

\begin{lstlisting}[language=C, caption={The program to exploit the vulnerability in \S\ref{sec:typeii-1}}, label={lst:typeii-1}]
v9 = syscall(SYS_open, (long)v8, 2, 0);
syscall(SYS_read, (long)v9, (long)v0, 5195);
syscall(SYS_unlink, (long)v3);
syscall(SYS_truncate, (long)v6, 4367);
syscall(SYS_unlink, (long)v7);
syscall(SYS_symlink, (long)v2, (long)v10);
syscall(SYS_lstat, (long)v2, (long)v1);
syscall(SYS_setxattr, (long)v2, (long)v12, (long)v11, 127, 1);
syscall(SYS_pread64, (long)v9, (long)v0, 6806, 299);
syscall(SYS_listxattr, (long)v10, (long)v1, 5210);
syscall(SYS_removexattr, (long)v4, (long)v13);
syscall(SYS_removexattr, (long)v2, (long)v14);
v15 = syscall(SYS_open, (long)v4, 2, 0);
syscall(SYS_listxattr, (long)v5, (long)v1, 5836);
syscall(SYS_utimes, (long)v5, (long)v1);
syscall(SYS_setxattr, (long)v2, (long)v17, (long)v16, 11, 1);
syscall(SYS_lstat, (long)v2, (long)v1);
syscall(SYS_pwrite64, (long)v9, (long)v1, 1772, 434);
\end{lstlisting}

To fix this bug, as we mentioned above, we should pay attention to the \texttt{NameLength} field.
The corresponding patch is shown in Listing~\ref{lst:patch-typeii}.

\begin{lstlisting}[language=C, caption={The patch to the vulnerability in \S\ref{sec:typeii-1}}, label={lst:patch-typeii}]
diff --git a/fs/ntfs3/record.c b/fs/ntfs3/record.c
index 66eb11e0965ef..a952cd7aa7a4b 100644
--- a/fs/ntfs3/record.c
+++ b/fs/ntfs3/record.c
@@ -265,6 +265,11 @@ struct ATTRIB *mi_enum_attr(struct mft_inode *mi, struct ATTRIB *attr)
            if (t16 + t32 > asize)
                return NULL;
    
+		if (attr->name_len &&
+		    le16_to_cpu(attr->name_off) + sizeof(short) * attr->name_len > t16) {
+			return NULL;
+		}
+
            return attr;
        }
\end{lstlisting}

\subsubsection{\texttt{4d42ecd}}
\label{sec:typeii-2}

\begin{figure}[t]
    \centering
    \includegraphics[width=0.4\columnwidth]{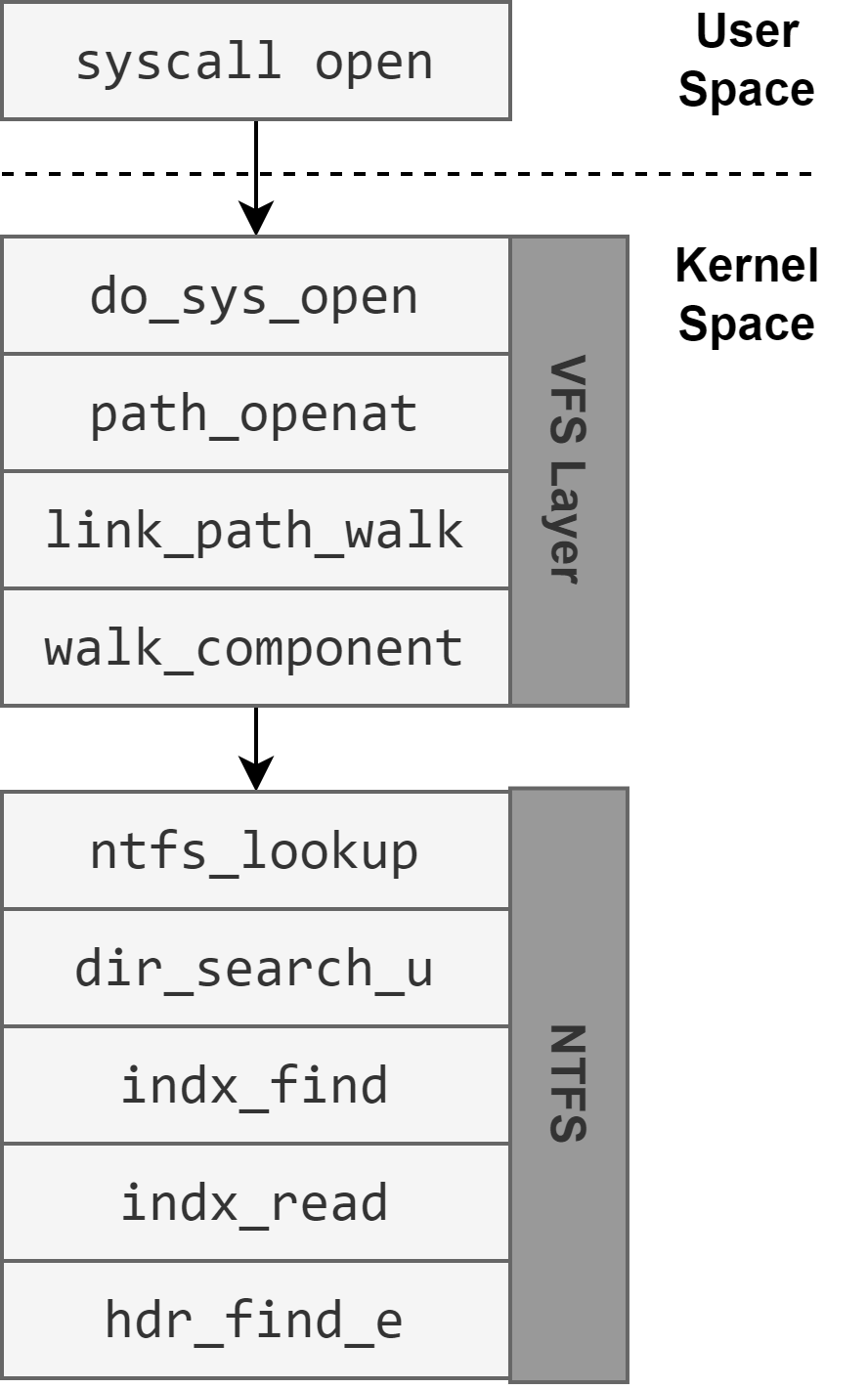}
    \caption{The trace of the exploitation on the bug in \S\ref{sec:typeii-2}.}
    \label{fig:typeii-2-trace}
\end{figure}

This bug is an out-of-bound read that is triggered by an \texttt{open} system call.
From the log file, we can conclude the trace as shown in Fig.~\ref{fig:typeii-2-trace}.
The bug is triggered in \texttt{hdr\_find\_e}.

There is a structure named index buffer in the NTFS disk image.
It is composed of a header and several entries.
The header is defined by a struct, named \texttt{INDEX\_BUFFER}, as shown in Listing~\ref{lst:index-buffer}.
The field \texttt{ihdr} holds some metadata of the buffer itself, like the length of the buffer and how many entries are used already.

\begin{lstlisting}[language=C, caption={A code snippet of \texttt{INDEX\_BUFFER} struct}, label={lst:index-buffer}]
struct INDEX_BUFFER {
	struct NTFS_RECORD_HEADER rhdr; // 'INDX'
	...
	struct INDEX_HDR ihdr; // stores metadata
};
\end{lstlisting}

The function \texttt{hdr\_find\_e} conducts a binary search in the index buffer to find a certain entry, i.e., a specific file index.
Note that the buffer is allocated by the kernel and its size is calculated from a variable read from disk, named as \texttt{index\_block\_size}. 
Specifically, the binary search adopts \texttt{ihdr->used} to calculate the end of the buffer.
If it's inconsistent with the result buffer size calculated from \texttt{index\_block\_size}, e.g., larger than the allocated size of index buffer, the binary search will access the outside.

The aim of the corresponding patch is to make sure the search cannot access the outside.
As we can see from Listing~\ref{lst:patch-typeii-2}, the \texttt{bytes} at L10 is the index buffer allocation size. \texttt{offsetof(struct INDEX\_BUFFER, ihdr) + ib->ihdr.used} calculates the index buffer size from \texttt{used}.
To this end, it guarantees the access should always be limited within the allocated buffer.

\begin{lstlisting}[language=C, caption={Patch to the bug in \S\ref{sec:typeii-2}}, label={lst:patch-typeii-2}]
diff --git a/fs/ntfs3/index.c b/fs/ntfs3/index.c
index 613036f9c6e66..bc656868cf8a8 100644
--- a/fs/ntfs3/index.c
+++ b/fs/ntfs3/index.c
@@ -1017,6 +1017,12 @@ ok:
            err = 0;
        }
    
+	/* check for index header length */
+	if (offsetof(struct INDEX_BUFFER, ihdr) + ib->ihdr.used > bytes) {
+		err = -EINVAL;
+		goto out;
+	}
+
        in->index = ib;
        *node = in;
\end{lstlisting}

\section{Lessons Learned}
\label{sec:lessons}

As we categorize the root causes of bugs identified by {\framework} in 
\S\ref{sec:expr_results:cause}, there are a couple of things we learned from those
findings, which we recommend file system developers to follow.

First of all, \textbf{user-controllable data should always be treated as untrusted input}.
When a file system image is mounted, the Linux VFS routes the \textit{mount} 
request to the corresponding file system handler which reads data from the disk 
and parses them in the memory.
If the file system image contains fields which would be used to derive an array 
index or memory pointer, those fields could be easily crafted to trigger 
out-of-bounds access in kernel space, which leads to system crash or even 
local privilege escalation.
Fortunately, the specifications of most file systems are well-documented.
A file system developer could follow the specification to strictly check every 
single chunk of data read from the disk.

Secondly, \textbf{type confusion issues should be paid more attention}.
In all Unix-like systems, an \texttt{inode} is used to describe a file, a 
directory, or other file system objects.
However, the handling logic to processing a file could be totally different to 
process a directory.
If an object is wrongly interpreted as another, unexpected behaviors occur.
Actually, programming languages without memory safety (e.g., C and C++) are 
prone to weaknesses in this type.
The Linux kernel is also evolving into an operating system with languages enforcing memory safety~\cite{linux-rust}.

Last but not least, \textbf{conducting a high code coverage fuzzing testing for file systems is necessary.}
To the best of our knowledge, no off-the-shelf fuzzer can efficiently fuzz a new 
file system in the Linux kernel.
Meanwhile, this also indicates that new file systems could be great targets for 
security researchers but not the best choices for users.
In particular, the \texttt{get\_tree()} handler of each file system (e.g., 
\texttt{ntfs\_fs\_get\_tree()} of NTFS3) would be a good entry point for testing 
funny file system images.
As illustrated in Fig.~\ref{fig:mount}, the functions to parse data retrieved 
from disk (e.g., \texttt{ntfs\_init\_from\_boot()}) may miss some important 
validation logic.
Moreover, each file system has a handler (e.g., \texttt{ntfs\_lookup()} of NTFS3) 
to search a file while handling an \texttt{open} system call (see Fig.~\ref{fig:typeii-2-trace}).
Security researchers could craft data related to the \texttt{lookup} 
procedure in a mutated disk image and see if that would cause out-of-bounds 
access.

\section{Related Work}
\label{sec:related}

Fuzzing has been proven effective in finding vulnerabilities in various softwares and kernel binaries, including file systems.
Vulnerabilities in file systems can be exploited due to two orthogonal root causes, which are often taken as targets for fuzzers, i.e., \textit{disk images} and \textit{file operation-specific system calls}.
Most existing fuzzers against file systems target either the former one~\cite{afl,fuzzbsd,schumilo2017kafl}, or the later one~\cite{syzkaller,trinity,triforce,xu2020krace}.
For example, kAFL implemented by S. Schumilo et al.~\cite{schumilo2017kafl} has improved the AFL~\cite{afl} to specifically target kernels. They have evaluated the kAFL across multiple operating systems by only mutating the disk images.
Contrarily, the \textsc{Krace} focuses on multi-thread vulnerabilities, which must be triggered by a certain sequence of file operations.
Some work~\cite{xu2019fuzzing,kim2019finding,kim2020finding,gross2022refuzz}, including {\framework}, takes both factors into consideration. However, {\framework} targets another critical but close-source file system, NTFS, and identifies a dozen of vulnerabilities that are acknowledged by the Linux kernel.

\section{Discussion}
\label{sec:discussion}

\noindent
\textbf{Unique Challenges in Fuzzing NTFS File Systems.}
Compared to fuzzing other file systems, fuzzing NTFS images has some unique challenges, which can be concluded mainly in twofold.
On the one hand, non-transparency in both terms of implementation and documentation hinders implementing a fuzzer.
There is no official implementation released, and its so-called official documentation lacks lots of technical details. Therefore, we have to manually compare multiple third-party releases and their corresponding documentations, and cross-reference which implementation is adopted by Microsoft.
On the other hand, such a non-transparency still occurs in checksum validation.
For example, the OEM field of the BPS should start with ``NTFS\ \ \ \ '' (NTFS + 4 spaces). However, NTFS still requires the bytes per sector must be greater than 512 bytes and be a power of 2. Such constraints are not well documented in any documentation and we have to manually dig them up from source code of third-party releases.

\noindent
\textbf{Necessity of Implementing the Parser.}
There were several NTFS parsers, e.g., NTFS-3G and Linux legacy NTFS. However, they are not sufficient for supporting a fuzzer like {\framework}.
Specifically, traditional parsers like NTFS-3G is responsible for validating the integrity of the given NTFS image. If something goes wrong, like a piece of problematic checksum, the parser will not parse and load the image at all.
Fuzzing, however, will constantly mutate the metadata of the image to try to figure out new bugs. Such deliberately introducing corrupted metadata will invalidate traditional parsers.
Therefore, the parser in {\framework} can not only parse the given image, but also automatically recover corrupted metadata, like recalculating checksums.
Moreover, it is lighter than traditional ones which are typically maintained for several years.
Therefore, the parser in {\framework} is not the reinvented wheel.

\noindent
\textbf{Advantages over Other Fuzzers.}
{\framework} has more advantages than other potential fuzzers against file systems.
For example, fuzzers specifically designed for file systems, like Janus and Hydra, cannot parse an NTFS image and conduct the following analysis.
Syzkaller could be usable in testing an NTFS image, because it mutates the program consisting of file operations to examine if any vulnerabilities can be triggered.
However, under the same environment as {\framework}, we ran Syzkaller, with advanced options (like KCOV and SyzLang) enabled, for 3 weeks and no valid results could be obtained.
We speculate that this phenomenon can be explained by the experimental results shown in Table~\ref{table:Trophy}.
For bugs we have identified, 10 out of 12 are due to buggy images. Only 2 of them can be triggered by a certain program, while it still requires a mutated disk as prerequisites.
Therefore, only composing a series of random file operations cannot effectively identify bugs embedded in NTFS images, which forces us to find alternative methods and develop {\framework}.

\section{Conclusion}
In summary, we have proposed a fuzzer, named {\framework}, specifically targeting NTFS images. We have released the two core components of {\framework}, i.e., the NTFS parser, and the LKL that has been ported to the latest Linux kernel with KASAN integrated.
Based on the efficiency and effectiveness of {\framework}, we have identified 3 assigned CVE 0-day vulnerabilities and 9 severe bugs within the latest release of the Linux kernel. All of them are confirmed by Linux maintainers and the corresponding patches of 9 out of them are merged into upstreams.
For these identified bugs and vulnerabilities, we have conducted a thorough empirical study including case studies on representative cases.
Finally, based on our investigations on those loopholes and exploits, we summarized a set of best practices for developers and security researchers.

\section*{Acknowledgment}
We specially thank all anonymous reviewers and our shepherd for their valuable suggestions that significantly improve the quality of this paper.

\newpage
\bibliographystyle{IEEEtran}
\bibliography{ref}

\end{document}